\documentclass{article}%
\usepackage{jheppub}
\usepackage{amsmath, amsfonts}
\usepackage{bm}
\usepackage{color}
\usepackage{amsmath}
\usepackage{amsfonts}
\usepackage{amssymb}
\usepackage{graphicx}%
\usepackage[utf8]{inputenc}
\usepackage{amsfonts}%
\usepackage{amsmath}%
\usepackage{amssymb}%
\usepackage{graphicx}
\usepackage{soul}
\usepackage{xcolor}
\pdfoutput=1

\title{Scalar correlators and normal modes in holographic neutron stars}
\author[a]{Tob\'\i as Canavesi}
\emailAdd{tcanavesi@fisica.unlp.edu.ar}
\affiliation[a]{Instituto de F\'{\i}sica La Plata - CONICET \\ C.C. 67. 1900 La Plata, Argentina 
}
\author[b]{Octavio Fierro}
\emailAdd{ofierro@ucsc.cl}
\affiliation[b]{Departamento de Matem\'{a}tica y F\'{\i}sica Aplicadas, Universidad
Cat\'{o}lica de la Sant\'{\i}sima Concepci\'{o}n,\\ Alonso de Ribera 2850,
Concepci\'{o}n, Chile.}
\author[a,c]{Nicol\'{a}s Grandi}
\emailAdd{grandi@fisica.unlp.edu.ar}
\affiliation[c]{Departamento de F\'{\i}sica - UNLP,\\ C.C. 67, 1900 La Plata, Argentina}
\author[a,c]{Pablo Pisani}
\emailAdd{pisani@fisica.unlp.edu.ar}

\abstract{We study a scalar field in the background of a holographic neutron star at finite temperature and analyze its asymptotic behavior to compute the two-point correlator at the boundary. From the normal modes of the field we determine the resonances and decay constants of the boundary field theory. We show that the correlator is dominated by the normal modes in the stable regions of phase space of the neutron star, becoming a power law as we move into the unstable zones.}

\begin{document}
\maketitle
\section{Introduction}

Since the original proposal by Maldacena, holography has become a new paradigm to study the physics of strong coupling in quantum field theory.

In particular, strongly coupled metallic systems have been studied in the holographic setup in the last decade, in relation to the strange metal phase experimentally reported on High $T_c$ superconductors \cite{keimer-2015,Chatterjee9346}. The first model to describe holographic metals was that of a spherical neutron star in asymptotically global AdS spacetime, known as {\it holographic neutron star} \cite{hns2010,hns2011}. Almost simultaneously, a somewhat more accurate description was proposed as a planar charged {\em electron star} that at infinity approaches AdS spacetime in the Poincar\'e patch \cite{es2011}. Since then, most of the research has focused on electron stars \cite{Hartnoll:2010xj,Hartnoll:2010ik,Hartnoll:2011dm,Puletti:2010de} and the  more standard holographic neutron star has received less attention.

However, the holographic neutron star constitutes an interesting  system on its own right. In studying it, the accumulated knowledge on spherical neutral solutions in asymptotically flat space  \cite{1990PhR...188..285P,Katz2003,2006IJMPB..20.3113C,2009PhR...480...57C,1999EPJC...11..173B,2015PhRvD..92l3527C,2019arXiv190810303C} can be adapted to the AdS setup. Since the strongly coupled holographic dual corresponds to a highly degenerate fermionic condensate on a conformal field theory defined on a sphere, this could be used to gain insight on the finite volume effects at strong coupling.

In previous works \cite{2018, 2020}, we explored the physics of the holographic neutron star at finite temperature. We learned that the solution space, spanned by the central temperature and the central degeneracy, is very rich: it includes configurations with a dense core and a diluted halo, as well as more regular non-cored solutions. We plotted the phase diagram of the bulk solutions and mapped the results into the boundary gauge theory. An interesting feature that we found is that, when the correlator of a scalar boundary operator is calculated in the geodesic approximation, it develops a {\em swallow tail} structure in some region of parameter space. We proved that such region corresponds to unstable configurations, defined by a dense cored profile and a diluted halo with a power-law edge.

In the present work, we explore further the holographic neutron star setup. Motivated by the previous observation that scalar correlators can be used as a proxy to detect instabilities, here we calculate the normal modes of a scalar operator, investigating their connection with the different features of the phase diagram.

The structure of this article is as follows. In Section \ref{review} we review the setup of the holographic neutron star. In Section \ref{scalar} we consider a scalar field on this background and present some typical profiles. The behavior of this field far away from the star gives the scalar correlators of the boundary theory. Section \ref{findings} summarizes our findings based on the numerical analysis of the asymptotics of the scalar field. To conclude, in Section \ref{outlook} we comment on two further aspects of this setting which will be explored in future work. An Appendix contains relevant results from pure global AdS which have been used for comparison with our model.

\section{The holographic neutron star at finite temperature}\label{review}

We work in the $3+1$ dimensional background of a {\em holographic neutron star}, which asymptotes global AdS spacetime, and is sourced in the interior by a large number of neutral self-gravitating fermionic particles.
The corresponding Einstein equations read
\begin{equation}
G_{\mu\nu}+\Lambda g_{\mu\nu}= 8\pi G\, T_{\mu\nu}\,,
\label{eq:einstein-crudas}
\end{equation}
where the cosmological constant $\Lambda$ can be written in terms of the AdS length as $\Lambda=-3/L^2$. The matter energy-momentum tensor is that of a perfect fluid
\begin{equation}
T_{\mu\nu}=P g_{\mu\nu}+(\rho+P)u_\mu u_\nu\,.
\label{eq:energy-momentum}
\end{equation}
We solve these equations with a stationary spherically symmetric Ansatz for the metric, with the form
\begin{equation}
ds^2=L^2\left(-e^{\nu(r)}\,dt^2+e^{\lambda(r)}\,dr^2+r^2\,d\Omega^2\right)\, ,
\label{eq:metricAdS}
\end{equation}
where $d\Omega^2=d \vartheta^2+\sin^2{\vartheta}d{\varphi}^2$. To write the pressure $P$ and density $\rho$, we assume that there is a large number of particles of rest mass ${m}_f$ within one AdS radius, or in other words we take the limit ${m}_fL\gg 1$  \cite{hns2010}. Then we can make the Thomas-Fermi approximation, writing
\begin{align}
\rho(r) &= \frac{g}{8\pi^3}\int f(r,\mathbf{p})\sqrt{\mathbf{p}^2+{m}^2_f }\,d^3\mathbf{p}\,,
\label{eq:rho}\\
P(r) &= \frac{g}{24\pi^3}\int
    f(r,\mathbf{p})\frac{\mathbf{p}^2}{\sqrt{\mathbf{p}^2+{m}^2_f }} \,d^3\mathbf{p}\,,
\label{eq:p}
\end{align}
where $g$ is the number of fermionic species. Here $f(r, \mathbf{p})$ is the Fermi-Dirac distribution function
\begin{equation}
f(r,\mathbf{p})=\frac1{e^{\frac{\sqrt{\mathbf{p}^2+ {m}^2_f}-\mu(r)}{T(r)}}+1} \,,
\label{eq:fermi-dirac-distribution}
\end{equation}
in terms of the local temperature  $T(r)$ and local chemical potential  $\mu(r)$, which are defined at equilibrium by Tolman and Klein conditions, respectively,
\begin{align}
e^{\nu(r)/2}\, T(r) &= T_0\,,
\label{eq:Tolman}\\
e^{\nu(r)/2}\, \mu(r) &= \mu_0\,,
\label{eq:Klein}
\end{align}
where we re-scaled the time variable such that $\nu(0)=0$, and then the constants $T_0=T(0)$ and \mbox{$\mu_0=\mu(0)$} represent the temperature and chemical potential at the centre of the configuration.
Replacing the Ansatz \eqref{eq:metricAdS}
into the Einstein equations (\ref{eq:einstein-crudas}) we obtain
\begin{equation}
e^{\lambda(r)}=\left(1-\frac{2GL^2 M(r)}{r}+r^2\right)^{-1}
\quad\quad
\mbox{with}\quad\quad
M(r)=4\pi\int_0^{r} \rho(\bar{r}) \bar{r}^2 d\bar{r}\,.
\label{eq:masa}
\end{equation}
The remaining equation for $\nu(r)$ reads
\begin{equation}
\frac{d\nu(r)}{dr}=\left(8\pi GL^2 P(r) +3\right)\,r\,e^{\lambda(r)}+\frac{\left(e^{\lambda(r)}-1\right)}{r}\,.
\end{equation}

According to the values of the central temperature $T_0$ and the central degeneracy ${\Theta_0=\mu_0/T_0-1}$, 
the density profiles resulting from the above dynamics manifest a well defined core-halo structure at large degeneracy and low temperature, that dissipates for lower degeneracies or larger temperatures \cite{2018, 2020}.

\begin{figure}[ht]
\centering
\vspace{-.6cm}
\includegraphics[width=1\textwidth]{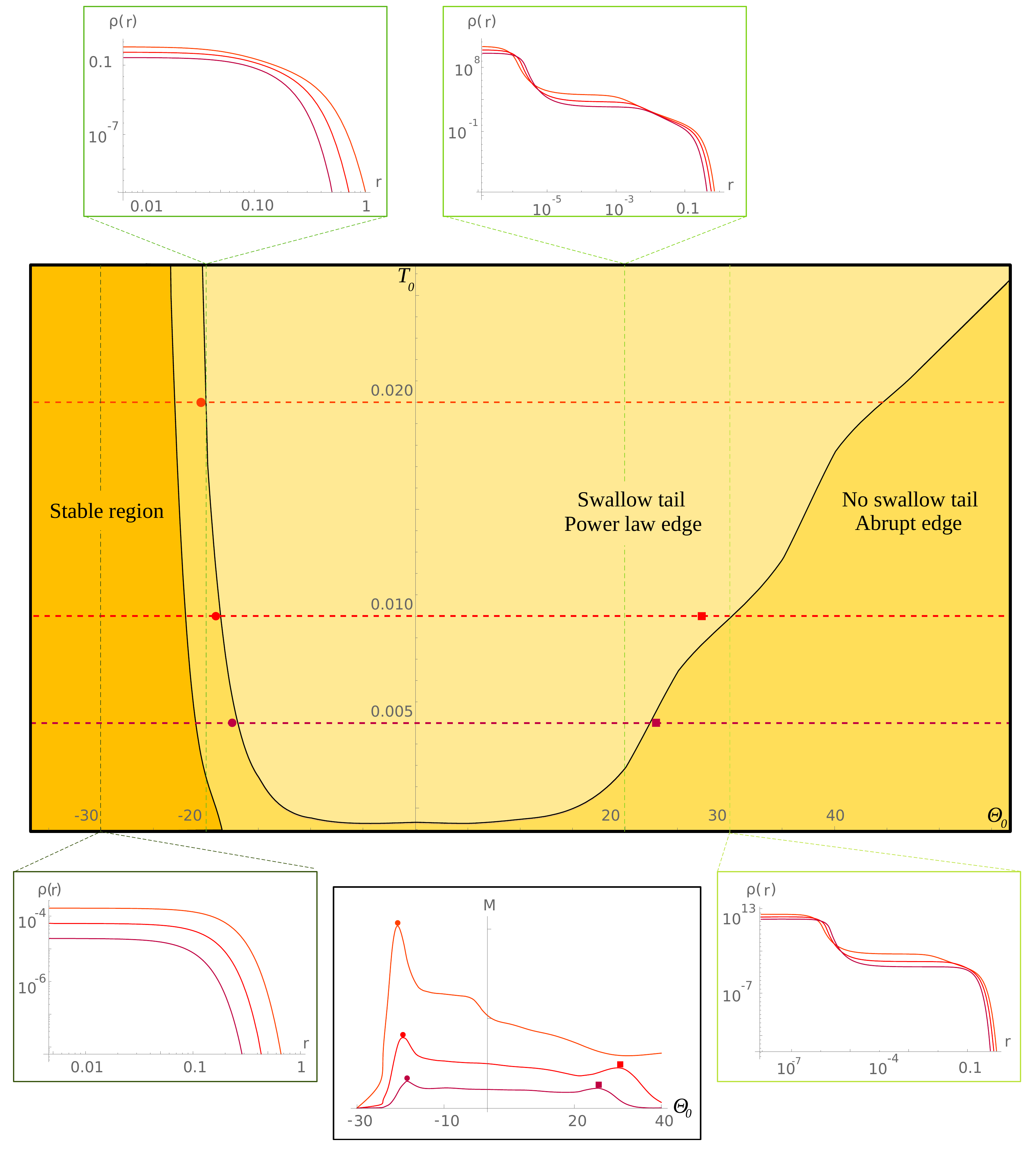}
\vspace{-1cm}
\caption{\label{fig:phasediagram}
Phase diagram of the holographic neutron star at finite temperature, in the central temperature $T_0$ {\em vs.} central degeneracy $\Theta_0$ plane. From left to right, the orange region correspond to a stable neutron star, which generically has a density profile with no special features and an abrupt edge. The intermediate yellow region denotes an unstable star with a single unstable mode, whose density profile has an abrupt edge. The light yellow region denotes an unstable star with more than one unstable mode, the density profile developing a power law dependence at its edge. As we move to the right for larger values of $\theta$ cored profiles appear, which according to the number of unstable modes may have a power-law edge. The dots and squares are correlated with the maxima of the total mass as a function of the central degeneracy, which can be seen in the central plot at the bottom of the figure.}
\end{figure}

The resulting profiles are thermodynamically stable as long as the central temperature and central degeneracy are small enough. Katz stability analysis shows that perturbative modes become unstable as those variables are increased \cite{2020}. When compared with the so-called turning point criterion, that identifies the instabilities as sitting at the turning points of the total mass as a function of the central parameters, we learn that the first unstable mode shows up prior to the turning point, which coincides with the appearance of a second unstable mode. Moreover, a second turning point coincides with the disappearance of such mode. See Fig.\ \ref{fig:phasediagram} for a phase diagram, with examples of the typical density profiles that can be found at any region, and a comparison with the turning point criterion.

\section{Scalar field perturbations}\label{scalar}

We linearly perturb the above background with a neutral scalar field. The dynamics for such scalar probe $\Phi$ is limited to the Klein-Gordon equation in the presence of the background solution  \eqref{eq:metricAdS}
\begin{equation}
\left(
{\Box-{\sf m}^2}\right)
{\Phi}=0\ .\label{eq}%
\end{equation}
This equation can be separated by writing ${\Phi}$ as a Fourier decomposition in time, and a superposition of spherical harmonics in the angular variables,
\begin{eqnarray}\nonumber
 {\Phi}\left(  t,r,\Omega\right)  &=&\sum_{\ell m }\int d\omega\ e^{-i\omega
t}R_{\omega \ell m}\left(  r\right)  Y_{\ell m}\left(  \Omega\right)
\\ &&
\qquad\qquad\qquad\mbox{with}\quad
Y_{\ell m}(\Omega)=
(-1)^m
\mbox{\scriptsize$
\sqrt{\frac{(2\ell+1)}{4\pi}\frac{(\ell-m)!}{(\ell+m)!}}
$\normalsize}
\,e^{i m \varphi}P_{\ell}^m(\cos\vartheta)
\ .\label{sep}
\end{eqnarray}
Here $\Omega=(\vartheta,\varphi)$ denotes the spherical angles and
$P_\ell^m(\cdot)$ an associated Legendre polynomial. 
The spherical harmonics fulfill the standard relation 
$\nabla_{S^{2}}^{2}Y_{\ell m}\left(  \Omega\right)  =-\ell\left( \ell+1\right)
Y_{\ell m}\left(  \Omega\right)$.

In our spherically symmetric background only the eigenvalue $\ell$ appears in the wave equation. The radial dependence then satisfies
\begin{equation}
  R^{\prime\prime}_{\omega \ell}
+
\frac{1}{2}
\left(
\nu^\prime-\lambda^\prime
+\frac {4}{r}
\right) R^{\prime}_{\omega \ell}
+
e^{\lambda}\left(  \omega^2 e^{-\nu}-\frac{\ell\left(
\ell\!+\!1\right)}{r^2} -L^2 {\sf m}^{2}\right)  R_{\omega \ell}=0\ ,\label{eq:R}%
\end{equation}
where we omitted the $m$ index in $R_{\omega \ell m}=:R_{\omega \ell}$ since it is evident from the equation that there will be no dependence on it.

The correct boundary conditions are obtained by expanding the equation at the extremes of the radial interval. First we go to large $r$ where the equation takes the form
\begin{equation}
  R^{\prime\prime}_{\omega \ell}
+
\frac {{4}}{r}
 R^{\prime}_{\omega \ell}
-
\frac{L^2{\sf m}^{2}}{r^2}
 R_{\omega \ell}=0\ ,\label{eq:Rlarger}%
\end{equation}
whose solution is
\begin{equation}
R_{\omega \ell}
=
a_{\omega\ell}(1+\dots) \,r^{-\Delta_-}
+
b_{\omega\ell} (1+\dots)\,r^{-\Delta_+}\ ,
\label{eq:conditionsInf}
\end{equation}
with $2\Delta_\pm={3\pm\sqrt{9+4\,{\sf m}^2L^2}}$ and the dots represent subleading negative integer powers of $r$. For our numerical calculations we choose ${\sf m}L=2$ which implies $\Delta_+=4$ and $\Delta_-=-1$. Notice that a normalizable mode would then require $a_{\omega\ell}=0$.
On the other hand, when we expand the equation at small values of the radius, we get
\begin{equation}
  R^{\prime\prime}_{\omega \ell}
+
\frac 2r
R^{\prime}_{\omega \ell}
-
\frac{\ell\left(
\ell\!+\!1\right)}{r^2}
R_{\omega \ell}=0\ .\label{eq:Rsmallr}%
\end{equation}
The solution of this is immediately
\begin{equation}
R_{\omega \ell}= A_{\omega\ell}\, r^\ell + B_{\omega\ell}\, r^{-\ell-1}.
\label{eq:asymp}
\end{equation}
A regular solution corresponds to $B_{\omega\ell}=0$.
This constraint results in the non-independence of the coefficients of the leading and subleading parts $a_{\omega\ell}$ and $b_{\omega\ell}$ at infinity, what would in turn quantize the values of $\omega$ for which a normal mode $a_{\omega\ell}=0$ is obtained.

\begin{figure}[ht]
	\centering
	\includegraphics[width=0.45\textwidth]{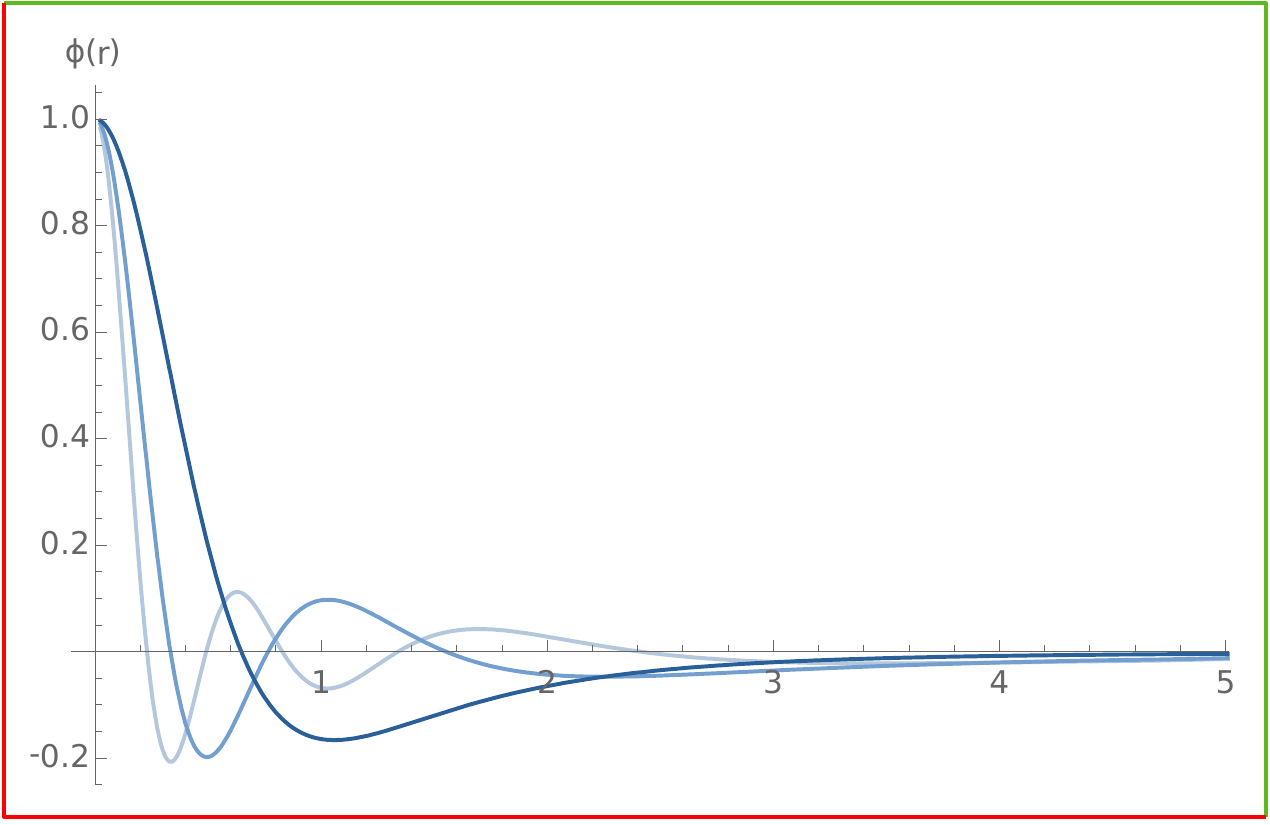}
	\includegraphics[width=0.45\textwidth]{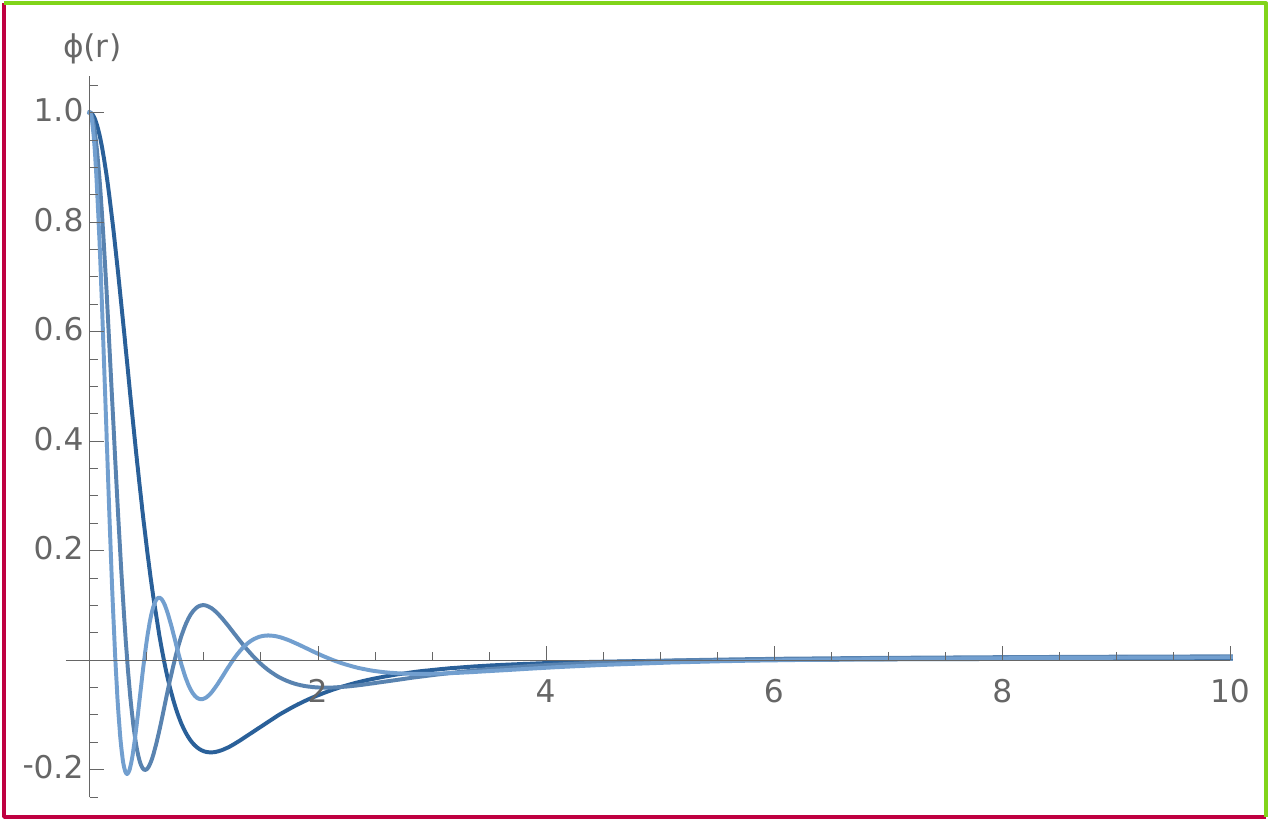}
	\caption{\label{fig:profiles} Different normal modes for $\tilde {T}_0=1/100$, $\theta_0=-20$ (Left) and $\tilde {T}_0=1/200$, $\theta_0=20$ (Right).}
\end{figure}

Typical profiles of the radial function are shown in Fig. \ref{fig:profiles}.

\section{Scalar correlators and normal modes}\label{findings}
The correlator of the dual scalar operator can be written as
\begin{equation}
\langle
{\cal O}(\Omega)
{\cal O}(0)
\rangle
=\sum_{\ell m} g_{\omega\ell}\,Y_{\ell m}(\Omega) \ ,
\label{eq:correlator}
\end{equation}
where use of the symmetries has been made to discard any dependence of $g_{\omega\ell}$ on $m$.  According to the holographic prescription, the coefficients $g_{\omega\ell}$ are given by the quotient of the subleading to the leading part of the dual bulk scalar field in \eqref{eq:conditionsInf}
\begin{equation}\label{gdeomegaele}
g_{\omega\ell} = \frac{b_{\omega\ell}}{a_{\omega\ell}}\ ,
\end{equation}

On general grounds, it is natural to expect that, as a function of $\omega$, the correlator $g_{\omega\ell}$ would have a set of simple poles plus an analytic contribution,
\begin{equation}
g_{\omega\ell}=\sum_{n}\frac{\rho_{n\ell}}{\omega-\omega_{n\ell}}+h_{\ell}(\omega)\, ,
\label{eq:fit}
\end{equation}
where $h_{\ell}(\omega)$ is analytic in $\omega$. The poles show up at the values of $\omega$ for which $a_{\omega\ell}=0$ in \eqref{eq:conditionsInf}. In other words, they correspond to the normalizable modes of the scalar field in the bulk, the so-called {\em normal modes}. The positions $\omega_{n\ell}$ of the poles are identified with the energy of the corresponding boundary excitations. On the other hand, the residua $\rho_{n\ell}$ correspond to their decay amplitudes.

\begin{figure}[h]
	\centering
	\includegraphics[width=0.5\textwidth]{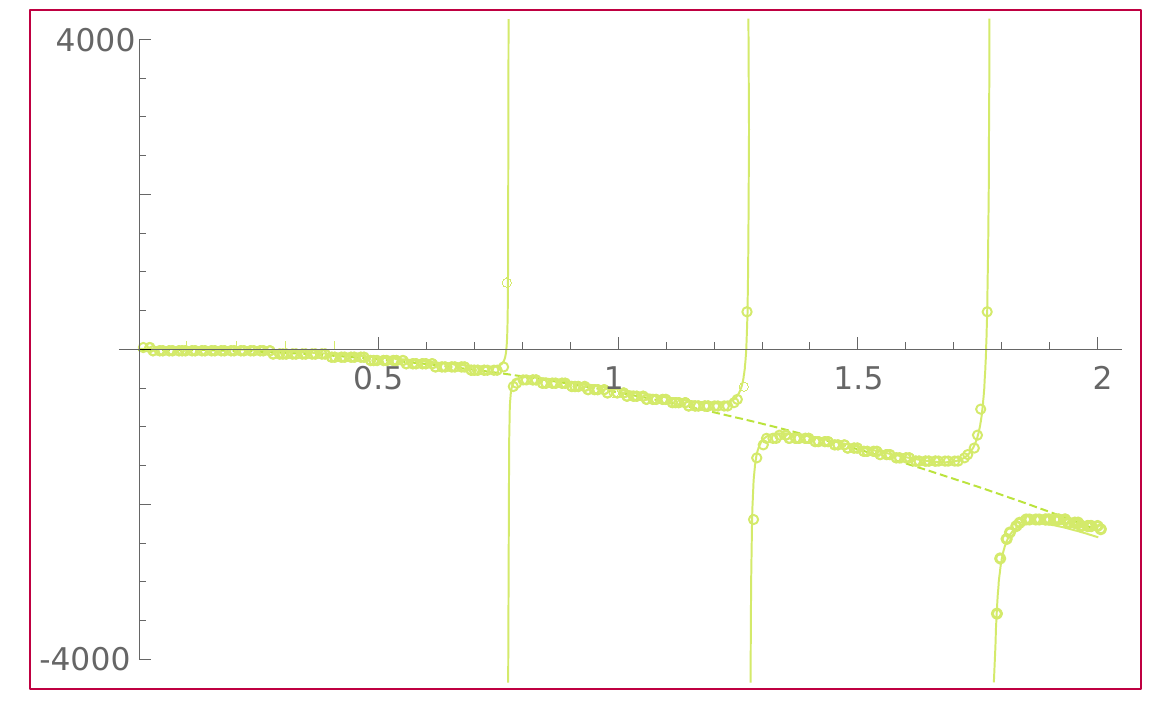}
	\caption{\label{fig:ajuste} Numerical form of the scalar correlators at $\Theta_0=30$ and $T_0=1/200$ for $\ell=0$, compared with the fit defined in equation \eqref{eq:fit} with a quadratic form for the analytic component $h_{\ell}(\omega)$.}
\end{figure}

In our numerical explorations of the parameter space, the expansion of the smooth part $h_{\ell}(\omega)$ up to the quadratic order in $\omega$ resulted in a good fit of the numerical data, see Fig.\ \ref{fig:ajuste}.

\begin{figure}[h]
	\centering
	\includegraphics[width=0.5\textwidth]{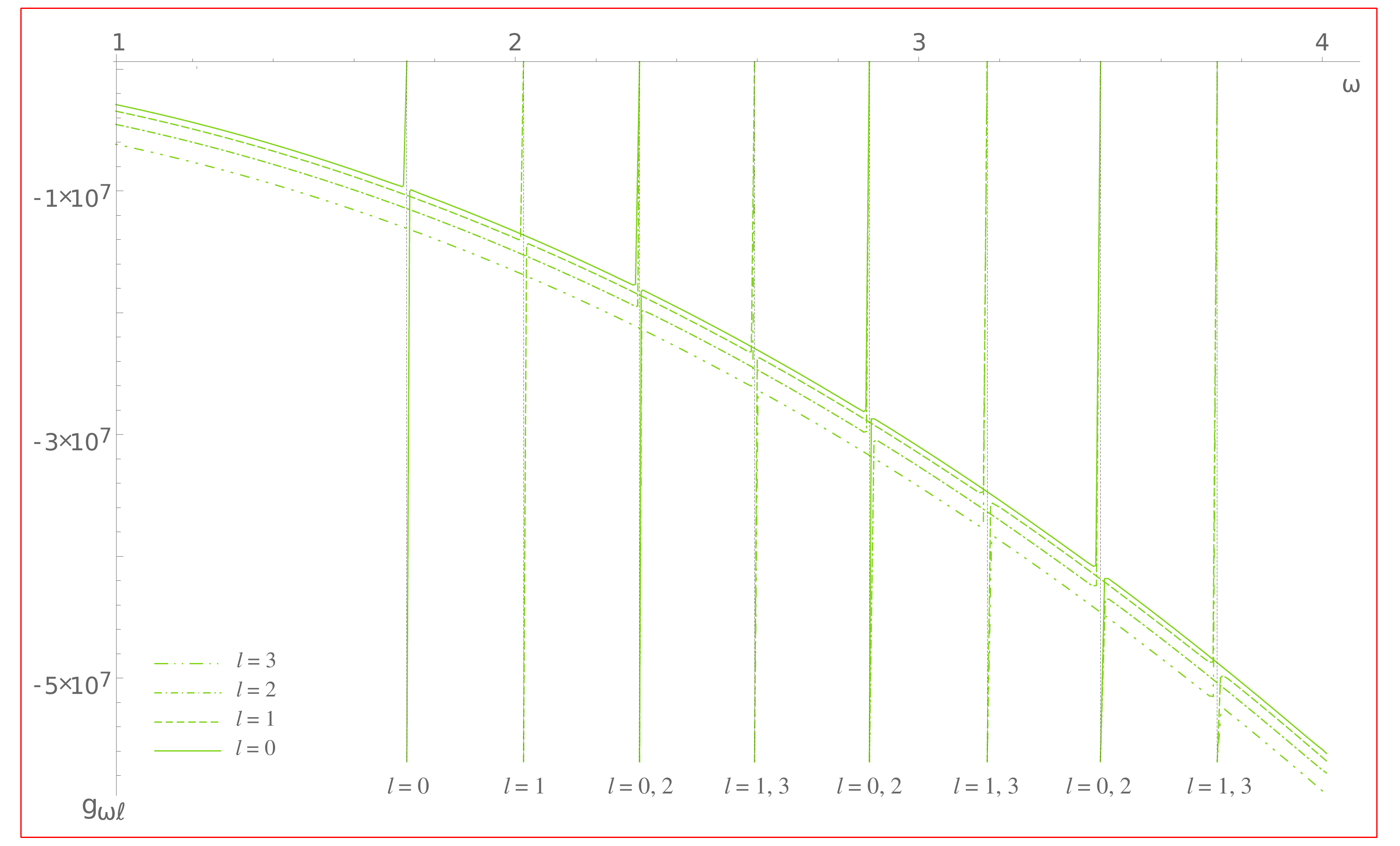}
	\caption{\label{fig:correlators2} Plots of the scalar correlators at $\Theta_0=20$ and $T_0=1/100$ for different values of $\ell$. To our numerical precision, the shift on the position of the poles as $\ell$ is increased is well approximated by the AdS expression $\omega_{n\,\ell\!+\!\Delta\ell}-\omega_{n\ell}=\Delta\ell$ 	.}
\end{figure}

As   can be seen in Fig. \ref{fig:correlators2}, as we increase $\ell$ the energies of the normal modes jump by an amount that, to our numerical precision, is well approximated by the AdS expression $\omega_{n\,\ell\!+\!\Delta\ell}-\omega_{n\ell}=\Delta\ell$ (see Appendix \ref{sec:appendixAdS}). However, as it could have been expected, the energy distance between different normal modes with the same angular momentum do not match the AdS formula $\omega_{n\!+\!\Delta n\,\ell}-\omega_{n\ell}\neq 2\Delta n$.

\begin{figure}[ht]
	\vspace{.15cm}
	\centering
	\includegraphics[width=0.95\textwidth]{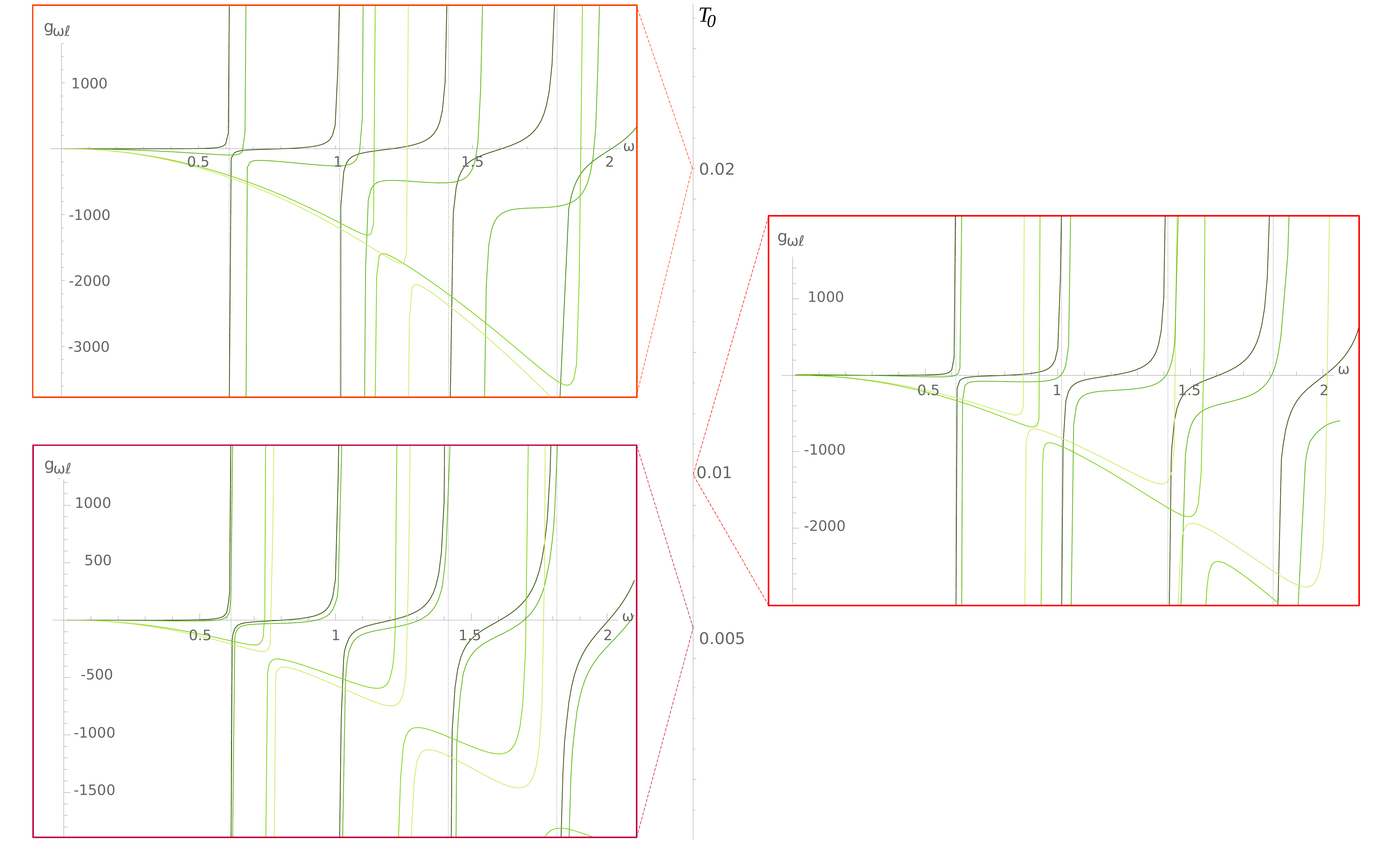}
	\caption{\label{fig:correlators}
	Plots of the scalar correlators. The frame color correlates with the different central temperatures $T_0$ depicted by the dotted horizontal lines in Fig. \ref{fig:phasediagram}, while that of the curves correspond to the different central degeneracies $\Theta_0$ depicted by the vertical dotted lines in the same figure.
	}
\end{figure}

The behaviour of the correlators as we change the central temperature $T_0$ and degeneracy $\Theta_0$ are shown in Fig.\ \ref{fig:correlators}. In the stable region of the phase diagram, the correlators look very much like those of the pure AdS case (see Appendix \ref{sec:appendixAdS}). As we increase $\Theta_0$ at fixed $T_0$, moving into the unstable region, the analytic part becomes more important, eventually dominating the plot. This is interpreted as an indication that the system is getting more and more critical, developing a power law correlator.

\begin{figure}[ht]
	\vspace{.01cm}
	\centering
	\includegraphics[width=0.43\textwidth]{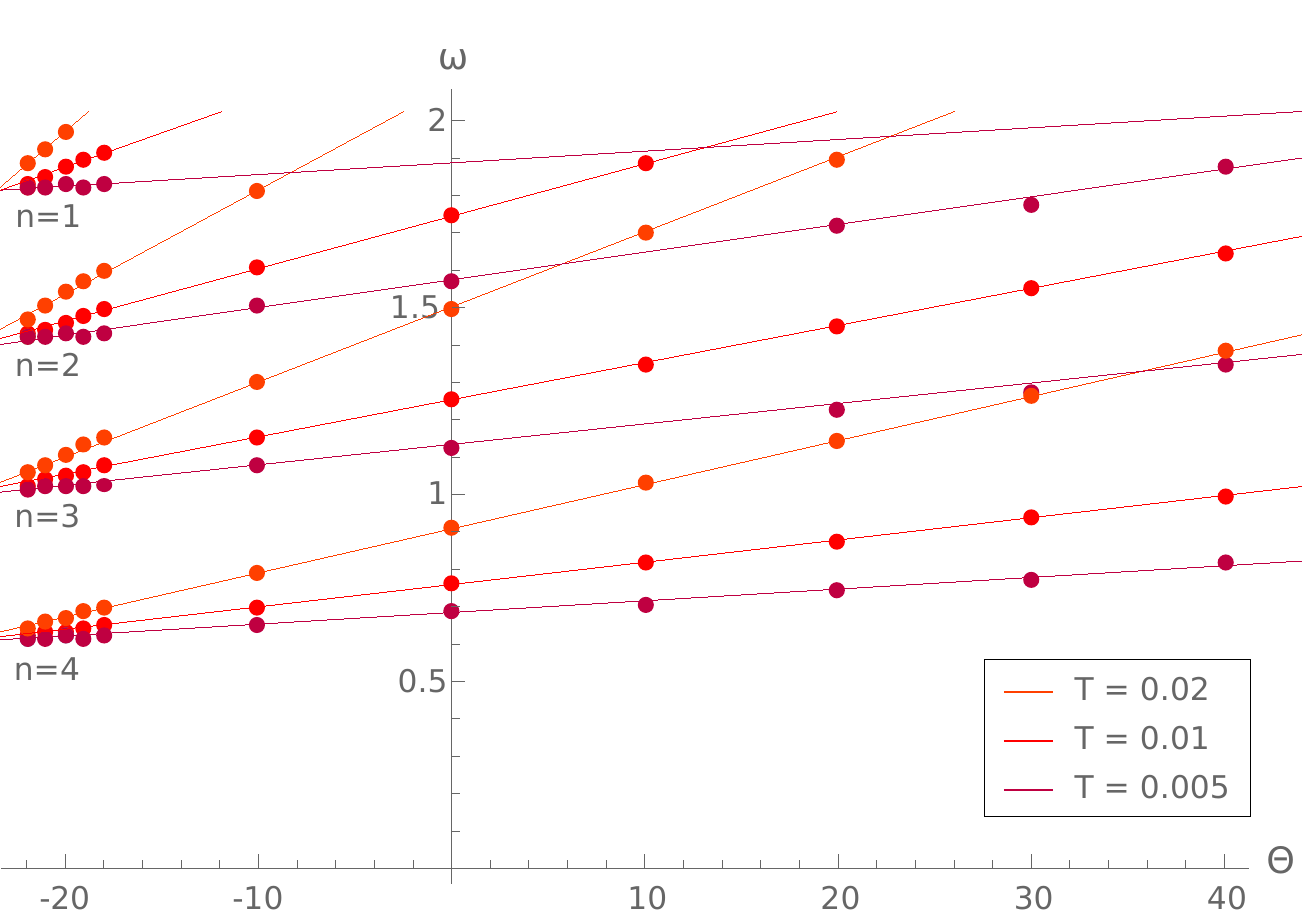}
	\hspace{0.05\textwidth}
	\includegraphics[width=0.43\textwidth]{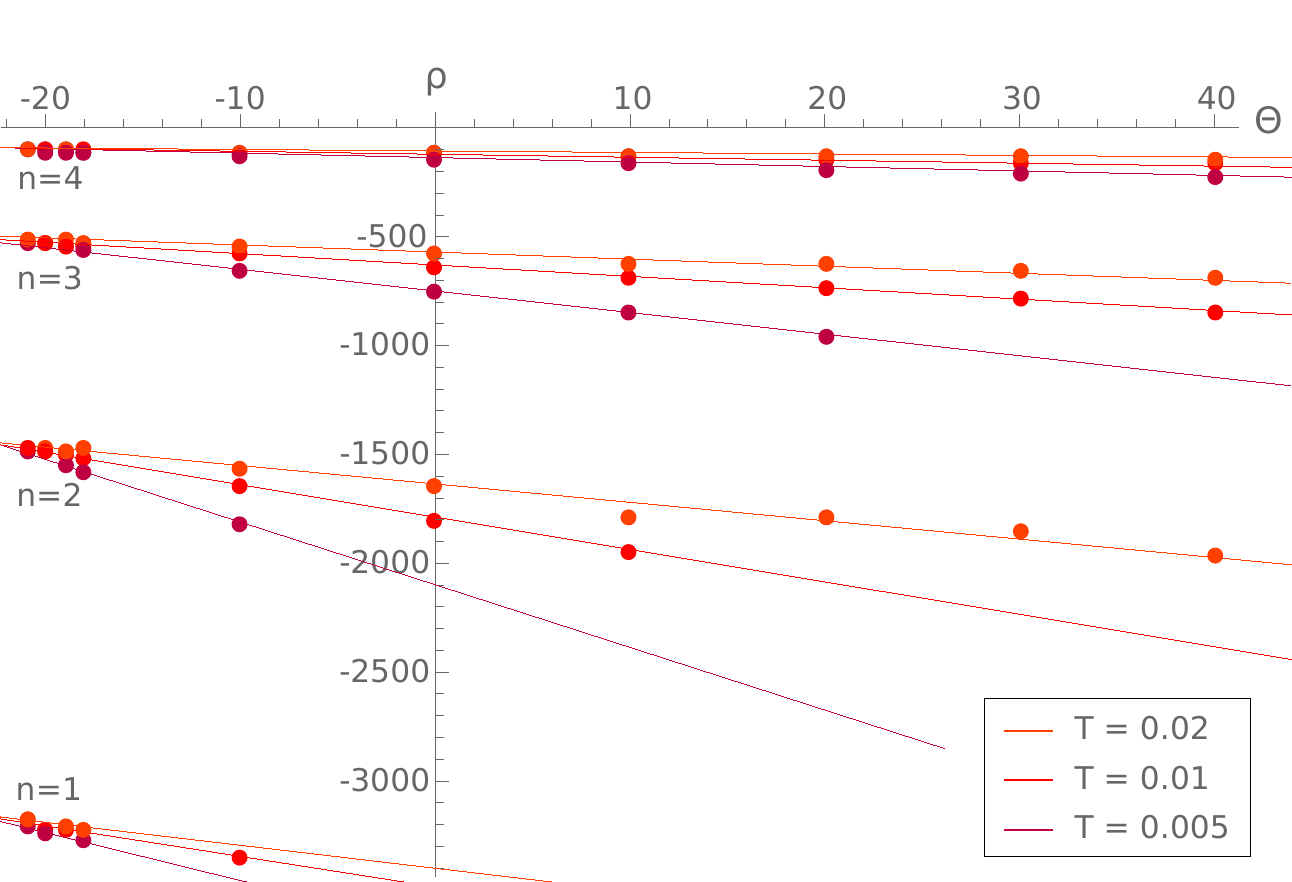}
	\caption{\label{fig:ajustes2} Plots of the normal modes energies (left) and decay constants (right) as functions fo the central degeneracies $\Theta_0$ for different central temperatures $T_0$. The shades of red correlate with the different central temperatures $T_0$ depicted by the dotted horizontal lines in Fig. \ref{fig:phasediagram}.}
\end{figure}

In the expression \eqref{eq:fit} the normal mode energy $\omega_{n\ell}$ and the absolute value of its decay constant $|\rho_{n\ell}|$ grow linearly both with the central temperature $T_0$ at fixed central degeneracy $\Theta_0$, and with  the central degeneracy $\Theta_0$ at fixed central temperature $T_0$, see Fig.\ \ref{fig:ajustes2}.

\section{Outlook}\label{outlook}

A first interesting question that is still open at this stage, is how does the swallow tail structure on the correlator, that was found on \cite{2018,2020} in the geodesic limit of large conformal dimensions $\Delta_+$, manifests itself in the present finite $\Delta_+$ context. In this article, the scalar correlators ---as defined in \eqref{eq:correlator} and \eqref{gdeomegaele}--- are obtained from the real frequency stationary states of a scalar field on a neutron star background with Lorentzian signature. These correlators show no new distinctive feature when entering the unstable region but display a smooth metamorphosis from a pole-dominated curve into a power law form. Nevertheless, we expect that the numerical analysis of classical solutions of the scalar field on the background of a Euclidean neutron star would provide a method to test whether the swallow tail also occurs at finite conformal dimensions.

In general, the correlator of a quantum field is divergent at coinciding points (in our case, vanishing angular separation $\Delta\vartheta=0$) but is expected to be smooth at antipodal points ({\em i.e.} at $\Delta\vartheta=\pi$).
For finite $\Delta_+$, a non-vanishing derivative at $\Delta\vartheta=\pi$ could be read from the coefficients of the partial wave expansion of the correlator. Due to the spherical symmetry of the bulk background, the correlator admits a Fourier expansion in terms of Legendre polynomials $P_\ell(\cos{\Delta\vartheta})$ with certain coefficients $C_\ell$.
An expansion of $C_\ell$ in powers of $\ell$ can be numerically studied. We checked that the leading behavior for large $\ell$ is $C_\ell\sim \ell^{2\Delta_+-1}$, as expected from the singularity at $\Delta \vartheta=0$. Positive powers of $\ell$ in the asymptotics of $C_\ell$ would lead to divergent series but, upon Borel summation, give finite contributions to the correlator. In particular, they are singular at $\Delta\vartheta=0$ but have vanishing derivative, and are thus smooth, at $\Delta\vartheta=\pi$.

The derivative of the correlator at $\Delta\vartheta=\pi$ can be read from the term in the asymptotics of $C_\ell$ which decreases as $\ell^a$ with $-3\leqslant a<-1$. A term with this type of falloff at large $\ell$, if it has no alternating sign, would give a non-vanishing contribution to the derivative. However, even for small conformal dimension, to determine such subleading behaviour we need to explore with higher numerical precision the large-$\ell$ behavior of the field; this task is to be addressed in a forthcoming work.

Another interesting question in the holographic neutron star system is the nature of the stable phase to which the system would decay when the instability is reached. In other words, that of the yellow regions in Fig.\ \ref{fig:phasediagram}. Such a phase would correspond to a bulk solution compatible with our boundary conditions, which contain in particular a nonvanishing chemical potential $\mu$. Since by Klein condition the chemical potential would remain non-vanishing at any interior point with a finite lapse function, the required bulk solution cannot be a black hole. Indeed, a black hole solution with our boundary condition for the chemical potential would have a non-vanishing chemical potential at the horizon. Thus, it would be singular, representing a non-equilibrium state. A viable alternative could be thermal AdS, since it has no horizon at which the non-vanishing chemical potential could become a problem. Another way out is to turn on additional degrees of freedom, in the form of additional bulk fields that vanish at the boundary, but deform the bulk allowing for a different phase. These are possibilities that are worth exploring.

\section{Acknowledgements}

The authors are grateful to Carlos Arg\"uelles, Diego Correa, Mart\'\i n Schvellinger and Guillermo Silva for discussion and helpful insights. This work is partially supported by CONICET grants PIP-2017-1109 and PUE084 ``Búsqueda de nueva física'', and UNLP grant PID-X791. This work was also partially supported by Chilean FONDECYT grant 11191175. O.F. would like to thank to the Direcci\'on de Investigaci\'on and Vicerrector\'ia de Investigaci\'on of the Universidad Cat\'olica de la Sant\'isima
Concepci\'on, Chile.

\newpage
\appendix
\section{Scalar normal modes in global AdS}\label{sec:appendixAdS}

In this appendix we determine the normal modes of a scalar field in global AdS$_4$,
\begin{align}
	ds^2
	=L^2\left\{
	-(1+r^2)\,dt^2+(1+r^2)^{-1}\,dr^2
	+r^2\,d\Omega^2\right\}\,.
\end{align}
The solutions of the equation of motion
\begin{align}
	\left(\Box-{\sf m}^2\right)\Phi(t,r,\Omega)=0\ ,
\end{align}
can be written as
\begin{align}
	\Phi(t,r,\Omega)=e^{-i\omega t}\,Y_{\ell m}(\Omega)\,R_{\omega\ell}(r)\,,
\end{align}
where the radial dependence is determined by
\begin{align}
	(1+r^2)\,R_{\omega\ell}''
	+\left(\frac2r+4r\right)\,R_{\omega\ell}'
	+\left(\frac{\omega^2}{1+r^2}
	-\frac{\ell(\ell+1)}{r^2}
	-L^2{\sf m}^2
	\right)R_{\omega\ell}=0\,.
\end{align}
Any solution to this equation which is regular at the origin is proportional to
\begin{align}
	R_{\omega\ell}(r)=
	\ \frac{r^\ell}{({1+r^2})^\frac{\omega}2}
	\ \mbox{}_2F_1\left(\tfrac{\ell+\Delta_-- \omega}2,\tfrac{\ell+\Delta_+- \omega}2;\ell+\tfrac32;-r^2\right)\,,
\end{align}
with
\begin{align}
	\Delta_\pm=\frac32\pm\sqrt{\frac94+{\sf m}^2}\,.
\end{align}
The behavior of the radial part at large $r$ can be read from the relation
\begin{align}\label{relhyper}
	\mbox{}_2F_1(a,b;c;-z)
	&=\tfrac{\Gamma(b-a)\Gamma(c)}{\Gamma(b)\Gamma(c-a)}
	\ z^{-a}
	\ \mbox{}_2F_1(a,a-c+1;a-b+1;-z^{-1})+\mbox{}\nonumber\\[2mm]
	&\mbox{}+\tfrac{\Gamma(a-b)\Gamma(c)}{\Gamma(a)\Gamma(c-b)}
	\ z^{-b}
	\ \mbox{}_2F_1(b,b-c+1;b-a+1;-z^{-1})\,.
\end{align}
Therefore, as $r\to\infty$,
\begin{align}\label{behatinf}
	R_{\omega\ell}(r)\sim
	a_{\omega\ell}\, r^{-\Delta_-}
	+b_{\omega\ell}\, r^{-\Delta_+}\,,
\end{align}
where
\begin{align}
	a_{\omega\ell}&=\tfrac{\Gamma\left(\frac12(\Delta_+-\Delta_-)\right)\,\Gamma\left(\ell+\frac32\right)}
	{\Gamma\left(\frac12(\ell+\Delta_+-\omega)\right)\,\Gamma\left(\frac12(\ell+\Delta_++\omega)\right)}\,,
	\label{A}\\[2mm]
	b_{\omega\ell}&=\tfrac{\Gamma\left(-\frac12(\Delta_+-\Delta_-)\right)\,\Gamma\left(\ell+\frac32\right)}
	{\Gamma\left(\frac12(\ell+\Delta_--\omega)\right)\,\Gamma\left(\frac12(\ell+\Delta_-+\omega)\right)}\,.
\end{align}
Notice that we have omitted the subleading contributions of the first term in \eqref{relhyper}.

If the mass parameter ${\sf m}$ is real and non-vanishing, then $\Delta_-<0$ and the leading behavior in \eqref{behatinf} diverges as $r\to\infty$, unless $a_{\omega\ell}=0$. This determines the oscillation frequencies $\omega_n$ of the normal modes:
\begin{align}
	\pm\omega_n=\Delta_++\ell+2n
	\qquad n=0,1,2,\ldots
\end{align}
\bibliography{biblio}

\end{document}